\documentclass[aps,prb,floatfix,superscriptaddress,twocolumn,footinbib]{revtex4-1}
\usepackage{amssymb}
\usepackage{amsmath}
\usepackage{amsfonts}
\usepackage{appendix}
\usepackage{bm}
\usepackage{graphicx}
\usepackage{epsfig}
\usepackage{epstopdf}
\usepackage{balance}
\usepackage[dvipsnames]{xcolor}
\usepackage{calc}
\usepackage{natbib}
\usepackage[colorlinks,
            linkcolor=blue,
            anchorcolor=blue,
            citecolor=blue,
            urlcolor=blue]{hyperref}
\usepackage{lipsum}
\usepackage{enumerate}

\begin{document}

\title{Gate-induced half-metals in Bernal stacked  graphene multilayer}

\author{Miao Liang}
\affiliation{School of Physics and Wuhan National High Magnetic Field Center,
Huazhong University of Science and Technology, Wuhan 430074,  China}

\author{Shuai Li}
\email{leeshuai@hust.edu.cn}
\affiliation{School of Physics and Wuhan National High Magnetic Field Center,
Huazhong University of Science and Technology, Wuhan 430074,  China}

\author{Jin-Hua Gao}
\email{jinhua@hust.edu.cn}
\affiliation{School of Physics and Wuhan National High Magnetic Field Center,
Huazhong University of Science and Technology, Wuhan 430074,  China}

\begin{abstract}
Recent experiments indicate that the Bernal stacked graphene multilayer (BGM) have an interaction induced gapped (or pseudo gapped) ground state.  Here, we propose  that, due to the electron correlation, the BGM can be induced into a half metallic phase  by applying a vertical electric field and  doping. The half metallic states in even-layer and odd-layer BGMs have  totally different behaviors, due to their different band structures. We systematically calculate the  graphene tetralayer (4L-BGM) and trilayer (3L-BGM) as the typical examples of the even-layer and odd-layer BGMs, respectively. 
In 4L-BGM, we find an interesting phenomenon of electric field induced inversion of the spin-polarized bands. Namely, in the half metallic phase, the spin polarization of the conducting channel and the net magnetic moment are inversed when the applied electric field exceeds a critical value. In 3L-BGM, a remarkable feature is that the inequivalence of the the two sublattices will intrinsically break the degeneracy of the the spin-up and spin-down bands  even in the zero electric field case. Our results suggest that 4L-BGM should be an ideal platform to detect the proposed half metallic phase in BGM systems.

\end{abstract}

\cite{}
\maketitle
\section{INTRODUCTION}
Graphene spintronics has draw great research interest in last twenty years\cite{ Tombros2007,2012Highly,spintronics2012,han_Gspin_2014,liu_spintronics_2020}. The main reason is that graphene has a very long spin diffusion length about several micrometers\cite{han_Gspin_2014,LSDL_PRL2016}, which makes it an ideal platform for spin device application. Manipulating the spin degree of freedom in graphene systems is not only of fundamental interest but also has potential device application\cite{RevModPhys.81.109,RevModPhys.83.407,Choudhuri2019,2015Science}. 

Half metal denotes a large family of materials in which electrons are conducting in one spin channel and insulating in the other\cite{groot1983,Halfmetal_F2008}. Thus, it is natural that the half metal materials are highly attractive for spintronics applications. The conventional half metal materials are mainly ferromagnets or ferrimagnets, \textit{e.g.}, Heussler alloys and transition metal oxides\cite{Halfmetal_F2008,Liu_2007,Yao_2005,Gao_2013,book_2016}. 

Interestingly, though pristine graphene is not a ferromagnetic material, it was proposed that a novel kind of correlated half metallic states can be realized in some graphene systems, namely graphene nanoribbon\cite{2006Half,2007Will,Kim_2008,2009Spin,Wu_2009}, graphene bilayer\cite{2013Possible} and rhombohedral stacked graphene multilayer (RGM)\cite{PhysRevB.98.075418}, by applying a proper external electric field and doping. Such novel half metallic state can be viewed as a new kind of half metallic antiferromagnets\cite{DEGROOT199145, PhysRevLett.74.1171, PhysRevLett.97.026401,2011HalfM}, which results from the interplay between the interaction induced antiferromagnetic spin order and external electric field.
Take the graphene bilayer as an example. Considering the Coulomb interaction, graphene bilayer can be described by Hubbard model. At half filling, one of the most possible ground states is the layer antiferromagnetic (LAF) state\cite{Zhangfan_2012,Stacking_2012,Honerkamp_2012,Lang_2012,LAFdft, Sun_2014,Rkov_2016}. The LAF state is a kind of spin density wave state, in which the spin polarization of two adjacent layers are opposite\cite{Otani_2010,Zhang_2011,Stacking_2012,PhysRevB.95.075422,Lee_2014}. When a perpendicular electric field is applied, spin-up and spin-down electrons feel different potential. So, the degeneracy of the spin up bands and spin down bands is broken. With a proper tiny doping, the doped electrons are filled into just one spin-polarized band, so that a half metallic ferromagnetic state appears. 
The case of the RGM is the same, since that it has a very similar interaction induced antiferromagnetic spin order (LAF state)\cite{Stacking_2012}. Actually, such half metallic state based on antiferromagnetic spin order is a general phenomenon in the Hubbard lattice models\cite{Garg_2014,Bag_2021}.  Note that the interaction induced half metallic phase in rhombohedral stacked graphene trilayer has been observed in recent experiments\cite{lee2019gate,2021Half}.

It is worth noting that the Bernal stacked graphene multilayers (BGMs), the most common form of graphene multilayer in nature,  also have correlated ground states (gapped or pseudogapped) at low temperature, as indicated by several transport experiments on suspended samples\cite{Nam_2016,0Insulating,2018A}. 
Basically, there are two kinds of graphene multilayers, \textit{i.e.}~the RGM and the BGM, according to the stacking order. Different stacking orders give rise to different band structures. Without regard to the electron-electron interaction, a N-layer BGM is metallic, which has N parabolic bands near the Fermi level for even N and an additional Dirac cone for odd N\cite{Guinea_2006,Latil_2006,Part_2007,Koshino_2007,Min_2008}. Although the density of states (DOS) of BGM at $E_f$ is not as large as that of the corresponding RGM, Coulomb interaction does induce a gap $\Delta_0$  at zero temperature for even-layer BGM. For example,  $\Delta_0 \approx 5.2$ meV for the tetralayer (4L-BGM) and $\Delta_0 \approx 13$ meV for the hexalayer (6L-BGM)\cite{2018A}. The odd-layer cases  have a similar phase transition as well, but the conductivity is always finite due to the additional linear bands\cite{2018A}. 

More importantly, a recent first principle calculations suggest that the correlated ground states of the BGMs also can host an antiferromagnetic spin order\cite{dft2020}, similar as  the LAF states in the RGMs.
Therefore, a natural question is that if the BGMs can be induced into a half metallic ferromagnetic state  via applying electric field and  doping.

In this work, based on the Hubbard model and self-consistent mean field method, we theoretically prove that it is possible to induce a half metallic phase in the BGMs by applying electric field and doping. The half metallic states in even-layer and odd-layer BGMs have rather different behaviors. For even-layer BGMs, we take the 4L-BGM as an example. Similar as the case of RGMs, a vertical electric field $E_0$ can break the degeneracy of the spin-up and spin-down bands, and a further doping is able to produce a half metallic phase with finite magnetic moment. Here, the spin-dependent gaps as  functions of $E_0$ here have unique behaviors different from the RGMs, which give rise to an interesting phenomenon of inversion of the spin-polarized bands. Such spin-polarized band inversion naturally defines a critical value of electric field $E_{inv}$. Once $E_0>E_{inv}$,  the spin polarization of conducting channel as well as the net magnetic moment of the half metal are inverted. The numerical results indicate that the required doping to achieve the half metallic states in BGM should be of the order of $10^{-10}cm^{-2}$, which is within the capability of the current experimental technique\cite{0Insulating}. 

We then use graphene trilayer (3L-BGM) as an example to illustrate the half metallic phase in odd-layer BGMs. In 3L-BGM, the half metallic state has some special features distinct from the even-layer BGMs.  First,  the degeneracy of the spin-up and spin-down bands are intrinsically broken without need of external electric field. It originates from the inequivalence of the two sublattices of  the odd-layer BGM, which thus should be a general feature of the Hubbard lattices with inequivalent sublattices. Second, the LAF state in 3L-BGM is rather robust against the perpendicular electric field, so that an extremely large $E_0$ is needed to destroy the spin order in 3L-BGM. However, $E_0$ can not obviously enlarge the split between the spin-up and spin-down bands in 3L-BGM. It implies that the required parameter range to realize the half metallic phase in 3L-BGM is much narrower than that in 4L-BGM. 

According to our calculations, we suggest that the 4L-BGM should be an ideal platform to detect the predicted half metallic phase of BGMs. A feasible way is to measure the spin-resolved gap (about several meV), considering the induced net magnetic moment is much smaller than that in RGM. Our work not only reveals a fundamental and interesting feature of the correlated ground states of BGM, but also enrich our understanding about the novel correlated half metal. 

The paper is organized as follow. In Section II, we give the model and methods used in the calculations. The calculated results and corresponding discussions are given in Section III. Finally, we give a summary in Section IV.  

\section{MODEL and Methods}

\begin{figure}[ht!]
\centering
\includegraphics[width=7.5cm]{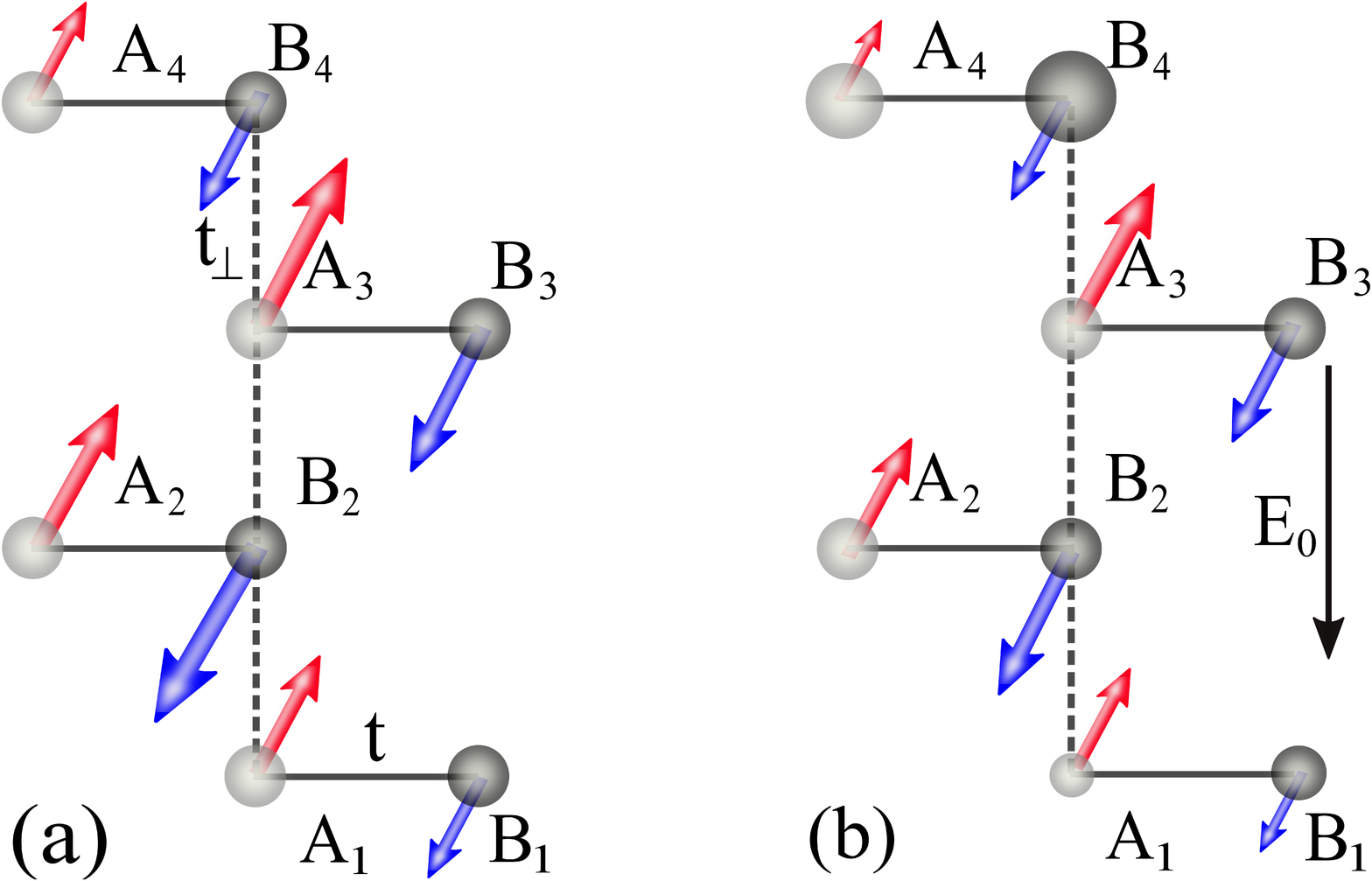}
\caption{(Color online) Schematic diagrams of charge (sphere) and spin (arrows) structures of 4L-BGM. (a) At half filling and in the absence of electric field. (b) At half filling and a vertical electric field $E_0$.
A and B denote the two sublattices of graphene, and subscripts $1-4$ are layer indices. The red (blue) arrows represent up-spin (down-spin). 
}
\label{fig1}
\end{figure}

We use Hubbard model and the self-consistent mean field method to calculate the correlated ground state of the BGM. Such methods have been successfully used to describe the interaction induced correlated states in graphene systems\cite{Nil_2006,Stacking_2012,2012Intrinsic,2013Possible}.  

The total Hamiltonian of a given BGM is \begin{equation}
\begin{aligned}\label{Htotal}
H=H_0+H_p+H_U.
\end{aligned}
\end{equation}
Here, $H_0=H_{intra}+H_{inter}$ is the tight binding Hamiltonian of the BGM, and $H_{intra}$ and $H_{inter}$ denote the intralayer and interlayer hopping, respectively. We only include  nearest-neighbor hopping in both $H_{intra}$ and $H_{inter}$, where
\begin{equation}
H_{\text {intra }}=-t \sum_{\left\langle li,lj\right\rangle \sigma}\left[a_{l \sigma}^{\dagger}(i) b_{l \sigma}(j)+\text { H.c. }\right]+\mu \sum_{l i \sigma} n_{l \sigma}(i),
\end{equation}
and 
\begin{equation}
H_{\text {inter }}=t_{\perp} \sum_{\left\langle li,l^{\prime}i^{\prime}\right\rangle \sigma}\left\{a_{l \sigma}^{\dagger}(i) b_{l^{\prime} \sigma}\left(i^{\prime}\right)+\mathrm{H.c.}\right\}.
\end{equation}
$a_{l \sigma}$ ($b_{l \sigma}$) is the annihilation operator of an electron on sublattice A (B).  $l$, $\sigma$, $i$ are the  index of layer, spin and site, respectively.  As shown in Fig.~\ref{fig1}, $t$ ($t_{\perp}$) is the intralayer (interlayer) nearest-neighbor hopping. $\left\langle li,li^{\prime}\right\rangle$ and $\left\langle li,l^{\prime}i^{\prime}\right\rangle$ represent the sum over all the intralayer and interlayer nearest-neighbor hopping, respectively. For BGM, we set $t=3.16$ eV and $t_\perp=0.381$ eV\cite{2013Possible}. $\mu$ is the chemical potential and $n_{l\sigma}(i)$ is the charge density on the sites.

 $H_p$ in Eq.~\eqref{Htotal} represents the electric field induced potential,
\begin{equation}\label{Hp}
H_{p}=\sum_{l i \sigma} \Delta_{l} n_{l \sigma}(i),
\end{equation}   
where $\Delta_l= -eV_l$ and $V_l$ is the potential of each layer in the presence of a vertical electric field $E_0$. Here, we consider the screening effect, which corresponds to  a charge redistribution between layers when  $E_0$ is applied.  
We define $E_l=(V_{l+1}-V_l)/d_0$ as the electric field  between two adjacent layers, where $l$ runs from 1 to $N-1$, and N is the number of layers of the system. $d_0=0.334 nm$ is the distance between two adjacent graphene layers in BGM. The graphene layers can be approximately viewed as conducting parallel plates. Electrostatics then tell us that 
\begin{equation}\label{El}
\begin{aligned}
E_l=E_0+2\pi{e}[\rho_l+\sum_{l^\prime}\rho_{l^\prime}(2\Theta(l-l^{\prime})-1)],
\end{aligned}
\end{equation}
where  $\rho_l=\sum_{\sigma,i}\left\langle n_{l\sigma}(i)\right\rangle/S$ is the electron density in each layer and  $S$ is the  area of one layer. $\Theta$ is the heavyside function. Therefore, once $\langle n_{l\sigma}(i)\rangle$ are given, we can determine  $V_l$ by calculating $E_l$ via Eq.~\eqref{El} , and thus get the value of $\Delta_l$ in $H_p$.  $\langle n_{l\sigma}(i)\rangle$  is calculated in a self-consistent way. And the midpoint of the BGM along  the perpendicular direction is  set to be the zero potential point. 

The Coulomb interaction is approximately described by the Hubbard U term . Using a  mean-field approximation, we have
\begin{equation}
H_{U}^{\mathrm{MF}}=U \sum_{l i \sigma}\left[\left\langle n_{l \sigma}(i)\right\rangle-1 / 2\right]\left[n_{l \bar{\sigma}}(i)-1 / 2\right],
\end{equation}
where $\bar{\sigma}=-\sigma$. 
$\left\langle n_{l \sigma}(i)\right\rangle$ has to be determined by the self-consistent method as well. The value of U is got by fitting the experimental value of correlation induced gap. 

We emphasize that in order to stabilize the antiferromagnetic spin order, an ultradense mesh of k points are needed in the self-consistent calculation. 

\begin{figure*}[ht!]
\centering
\includegraphics[width=14cm]{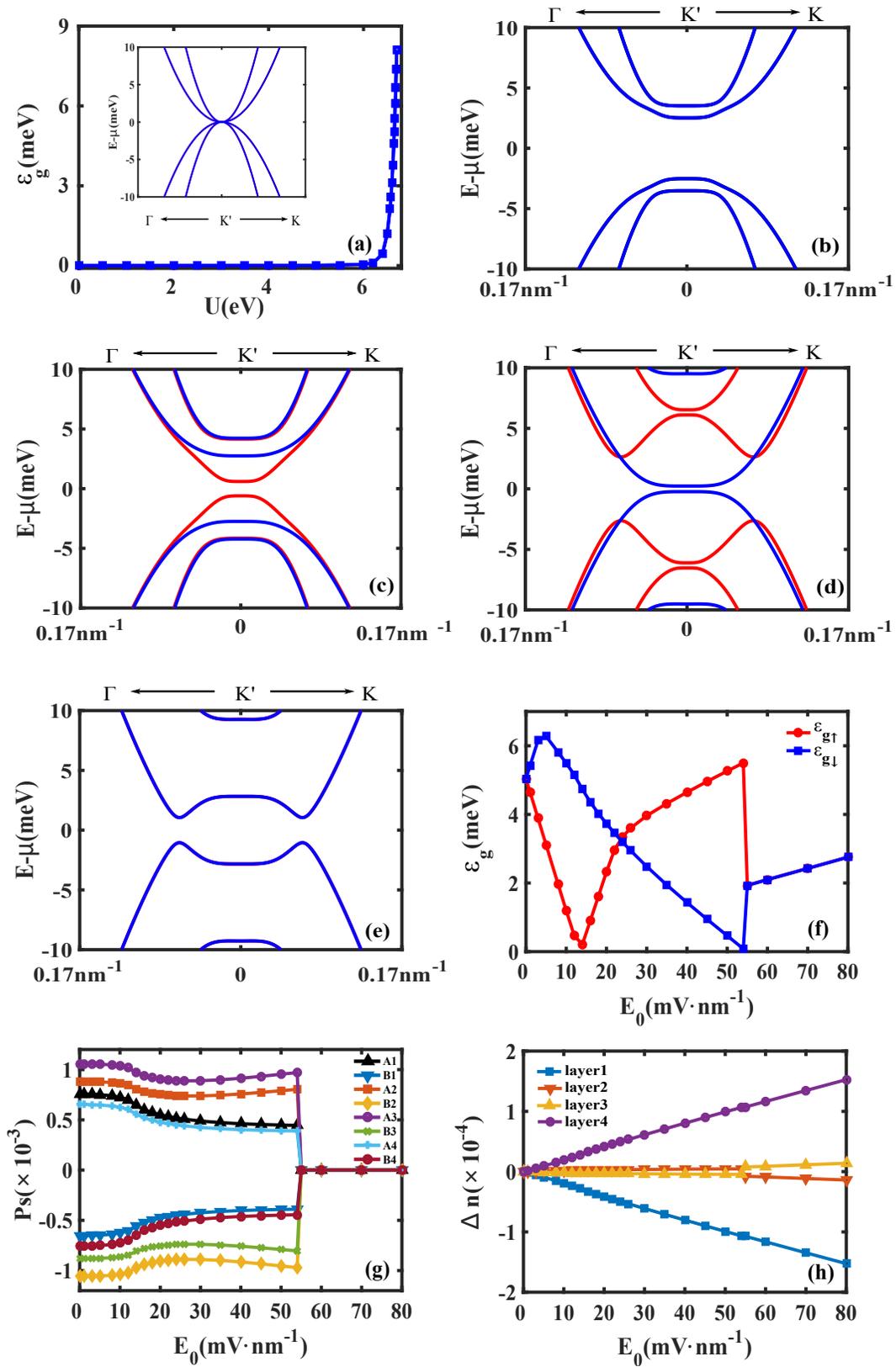}
\caption{(Color online) 4L-BGM at half filling. (a) Energy gap of 4L-BGM as a function of $U$ and in the absence of $E_0$. Insets: low-energy bands with $U=0$. (b) Low-energy bands with $U=6.65$ $eV$, $t=3.16$ $eV$, and $t_{\bot}=0.381$ $eV$. The following (c), (d) and (e) use the same parameters as that in (b). (c) and (d) show energy bands of the LAF stateS at (c) $E_0=10$ $mVnm^{-1}$ and (d) $E_0=50$ $mVnm^{-1}$, respectively. Blue (red) solid lines denote energy bands for spin-up (down) channel. (e) Energy bands of the LCP state in 4L-BGM with $E_0=60$ $mVnm^{-1}$. (f) Spin-resolved energy gaps against electric field $E_0$. (g) Spin polarization $P_s$ at each site in the unit cell as a function of $E_0$. (h) The charge variation $\Delta n$ on each layer as a function of $E_0$.
}
\label{fig2}  
\end{figure*}

\section{RESULTS AND DISCUSSIONS}

\subsection{Half metallic state in 4L-BGM}
Taking the 4L-BGM as a typical example, we first discuss the half metal states in the even-layer BGM. The calculated results are shown in Fig.~\ref{fig2}.

In single particle picture, the 4L-BGM has two parabolic conduction bands and two parabolic valence bands touching at the Fermi level, as shown in the inset of Fig.~\ref{fig2} (a). When the electron-electron interaction is considered,  our calculation suggests that the on-site Coulomb repulsion can induce a gapped SDW-like ground state with antiferromagnetic spin order. It is in good agreement with the recent DFT calculation\cite{dft2020}. The corresponding spin structure is shown in Fig.~\ref{fig1} (a), where the two sublattices (A and B) have opposite spin polarization.  Since the two sublattices are equivalent in 4LG-BGM, the spin-up and spin-down bands are degenerate, as shown in Fig.~\ref{fig2} (b). The interaction induced gap $\epsilon_g$ depends on the value of $U$. As illustrated in Fig.~\ref{fig2} (a), the larger $U$, the larger the gap. In our calculation, we use $U=6.65$ eV in order to fit the  experimental value of the gap of 4L-BGM, \textit{i.e.}, $\Delta_0 \approx 5.2$ meV at zero temperature\cite{2018A}. 

Such correlated ground state can be viewed as a kind of LAF state. First, it has antiferromagnetic spin order, see in Fig.~\ref{fig1} (a), and the net spin polarization of the whole system is zero. Meanwhile, the net spin polarization in each layer is nonzero, \textit{i.e.}, ferrimagnetism. And that of the two adjacent layers are opposite. 
These features are the same as the predicted LAF states in graphene bilayer and RGM. However, the LAF state of the BGM has its own unique features. In 4LG-BGM, the net spin polarization mainly distribute in the inner two layers, while in RGM the maximum value of the spin polarization occurs at the outermost two layers\cite{Stacking_2012}. 

We then discuss the effect of  electric field in the half-filled case.
An important fact is that the distributions of spin up and down electrons along the perpendicular direction do not coincide, as illustrated in Fig.~\ref{fig1} (a). 
Thus, when an vertical electric field is applied, up spin and down spin feel different potential, and the degeneracy of the spin up and down bands is lifted. 
In Fig.~\ref{fig2} (c), we see that an electric field $E_0=10$ $mVnm^{-1}$ obviously decreases the gap of spin-up bands (red solid lines), while that of the  spin-down band has a tiny change (blue solid lines). Namely, the conduction and valence bands now are all spin-up bands. The corresponding spin and charge distributions are illustrated in Fig.~\ref{fig1} (b).
We see that the spin and charge distributions are slightly changed by the vertical electric field, but it remains in the LAF state.
Very interestingly, when we use a larger electric field $E_0=50$ $mVnm^{-1}$ as in Fig.~\ref{fig2} (d), the situation reverses. The gap of  spin-down bands  (blue solid lines) is now suppressed instead, and becomes smaller than that of the  spin-up bands. 
Meanwhile, due to the electric field effect, the spin-up bands change saliently from the parabolic dispersion to the Mexican hat shape.
Such unique phenomenon implies there exists an inversion of spin-polarized bands in 4L-BGM when increasing $E_0$.  Further increasing $E_0$, a transition from the LAF state to the layered charge polarized (LCP) state \cite{Zhangfan_2012,2013Possible} occurs at the critical value  $E_c=54$ $mV nm^{-1}$. Then, the bands become spin degenerate, see in Fig.~\ref{fig2} (e) with $E_0=60$ $mVnm^{-1}$.

\begin{figure*}[ht]
\centering
\includegraphics[width=14cm]{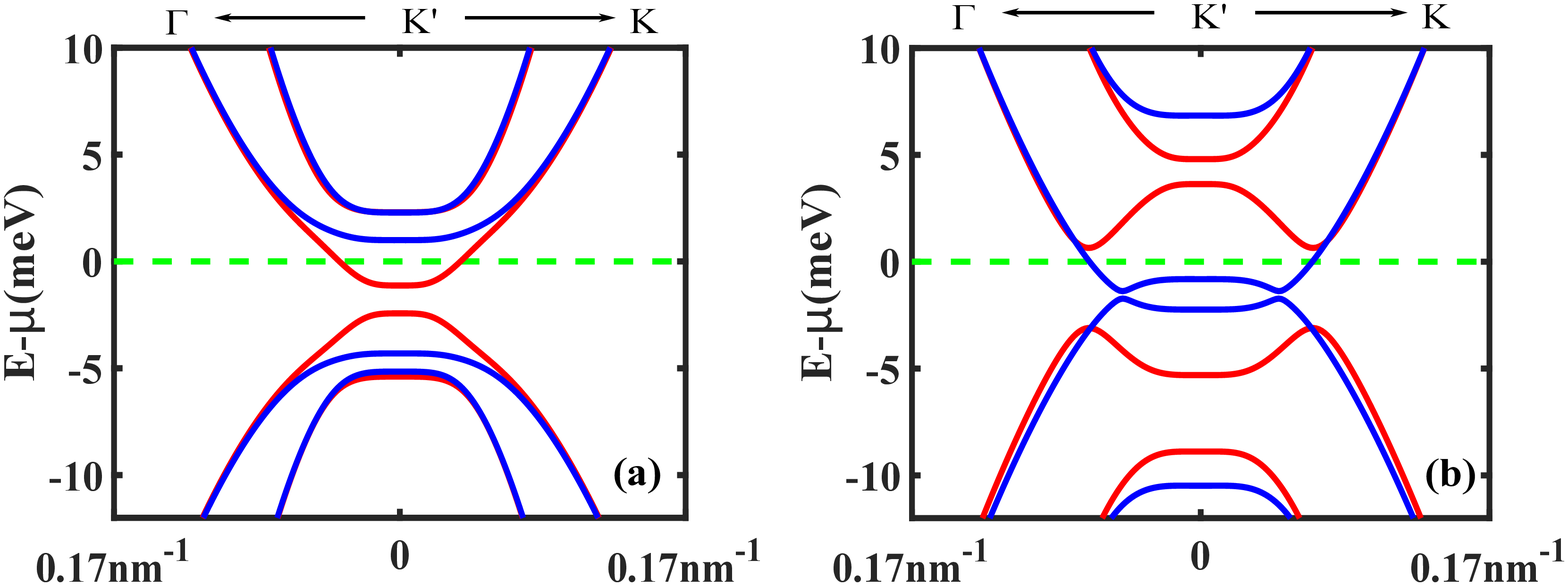}
\caption{(Color online) Half metallic phases in 4L-BGM. (a) Energy bands of the half metallic phase with a up-polarized net magnetic moment. The charge density is $\delta n\approx2.21\times10^{10}$ $cm^{-2}$ and other parameters are the same as that in Fig.~\ref{fig2} (c). (b) Energy bands of the half-metallic state with a down-polarized net magnetic moment. The charge density is $\delta n\approx6.28\times10^{10}$ $cm^{-2}$ and the parameters used are the same as Fig.~\ref{fig2} (d). The Fermi level $E_f$ is set at zero energy and is represented by the green horizontal dashed lines.
}
\label{fig3}
\end{figure*}

The  spin-polarized band inversion is shown more clearly in Fig.~\ref{fig2} (f), where we plot both  $\epsilon_{g_{\uparrow}}$ (gap of spin-up bands, red lines) and $\epsilon_{g_{\downarrow}}$ (gap of spin-down bands, blue lines) as  functions of  $E_0$. Increasing $E_0$, the  $\epsilon_{g_{\uparrow}}$ decreases at first and then increases, while the $\epsilon_{g_{\downarrow}}$ has the opposite behavior. The two curves  have an intersection point at $E_{inv}\approx24$ $mVnm^{-1}$, where $\epsilon_{g_{\uparrow}}=\epsilon_{g_{\downarrow}}$. So, $E_0$ actually has three different regions:

\begin{enumerate}[(1)]
\item  $E_0<E_{inv}$. In this region,  $\epsilon_{g_{\uparrow}}<\epsilon_{g_{\downarrow}}$.  $\epsilon_{g_{\downarrow}}$ first increases to a maximal value about $6.3$ meV and then decreases, while the $\epsilon_{g_{\uparrow}}$ first decreases to a minimal value  about $0.2$ meV and then increases.  The corresponding energy bands are like that in Fig.~\ref{fig2} (c). Because that both the valence and conduction bands are spin-up bands in this situation, it is in some sense can be viewed as a half semiconductor.
  
\item $E_{inv}<E_0<E_c$.  In this region,  $\epsilon_{g_{\downarrow}}<\epsilon_{g_{\uparrow}}$. The energy bands in this region are like that in Fig.~\ref{fig2} (d). Now, it is still in the half semiconductor state, but both the valence and conduction bands are spin-down bands. Note that  the spin-up  bands have been changed  to Mexican hat shape.

\item $E_0>E_c$. In this region, the bands become spin degenerate, see in Fig.~\ref{fig2} (e). It indicates a phase transition from the LAF to LCP state. As $E_0$ increases further, the gap increases monotonously. 
\end{enumerate}

We emphasize that the electric field induced spin-polarized band inversion is an unique feature of the 4L-BGM, which is distinct from that in graphene bilayer or RGM. In graphene bilayer, as $E_0$ increases, the gap of one spin increases  and that of the other spin decreases until $E_0$ exceed a critical value and the phase transition occurs\cite{2013Possible}. Namely, there is no such spin-polarized band inversion in graphene bilayer and RGM. 

In Fig.~\ref{fig2} (g), we plot the spin polarization $P_s$ on each site as a function of $E_0$, in which the phase transition at $E_c$ is shown very clearly. We also plot the charge variation $\Delta n$ on each layer as a function of $E_0$ in Fig.~\ref{fig2} (h). It indicates that, as $E_0$ increases, an obvious charge transfer from the bottom layer (layer 1) to the top layer (layer 4) occurs. Meanwhile, the charge on the middle two layer are nearly invariant. 

In the cases of graphene bilayer and RGM, we can get a half-metal state by slightly doping the LAF state in the presence of $E_0$.  The similar mechanism is also valid for the 4L-BGM.  A tiny doping can shift the Fermi level and partially fill the spin-polarized conduction (or valence) band of LAF state, which gives rise to a half metallic phase. However, as mentioned above,  4L-BGM has two different LAF states with different $E_0$, where the spin polarization of the conduction and valence bands are opposite. Therefore, 4L-BGM has two half metal regions, as shown in Fig.~\ref{fig3}, whose net magnetic moments have opposite direction. In Fig.~\ref{fig3} (a), the  charge density is about $\delta n\approx2.21\times10^{10} cm^{-2}$ with $E_0=10$ $mVnm^{-1}$, while the  charge density is $\delta n\approx6.28\times10^{10} cm^{-2}$  in Fig.~\ref{fig3} (b) with $E_0=50$ $mVnm^{-1}$. 
Apparently, the conducting channels in Fig.~\ref{fig3} (a) and (b) have opposite spin polarization. 
A larger doping will directly destroy the LAF states, and thus kill the half metallic phase.  For example, a doping about $\delta n\approx1.7\times10^{11} cm^{-2}$ is large enough to kill the half metallic phase of 4L-BGM. 

The induced magnetic moment and the spin-resolved gap are two markers of the half metallic phase in 4L-BGM. Since the net magnetic moment is zero at half filling, only the doped electrons contribute the net magnetic moment, see in Fig.~\ref{fig3}. So, the magnetic moment density of the half metallic phase in 4L-BGM should be of the order of $10^{10}\mu_B$ $cm^{-2}$. Meanwhile, we argue that transport experiments to measure the spin-resolved gap (about several meV as shown in Fig.~\ref{fig2} and Fig.~\ref{fig3}) is also a feasible way to identify the half metallic phase. Considering that the correlated half metallic phase in rhombohedral stacked graphene trilayer has been observed in experiment very recently\cite{lee2019gate,2021Half}, we expect that the proposed half metallic phase in 4L-BGM can be confirmed in future experiments. 

\begin{figure}[h]
\centering
\includegraphics[width=7.5cm]{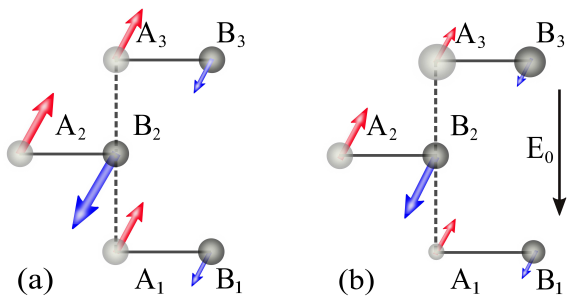}
\caption{(Color online) Schematic illustration of the charge (sphere) and spin(arrows) strutures of 3L-BGM. (a) At half-filling and in the absence of electric field. (b) At half-filling and a finite perpendicular electric field $E_0$.  
}
\label{fig4}
\end{figure}

\begin{figure*}[ht]
\centering
\includegraphics[width=14cm]{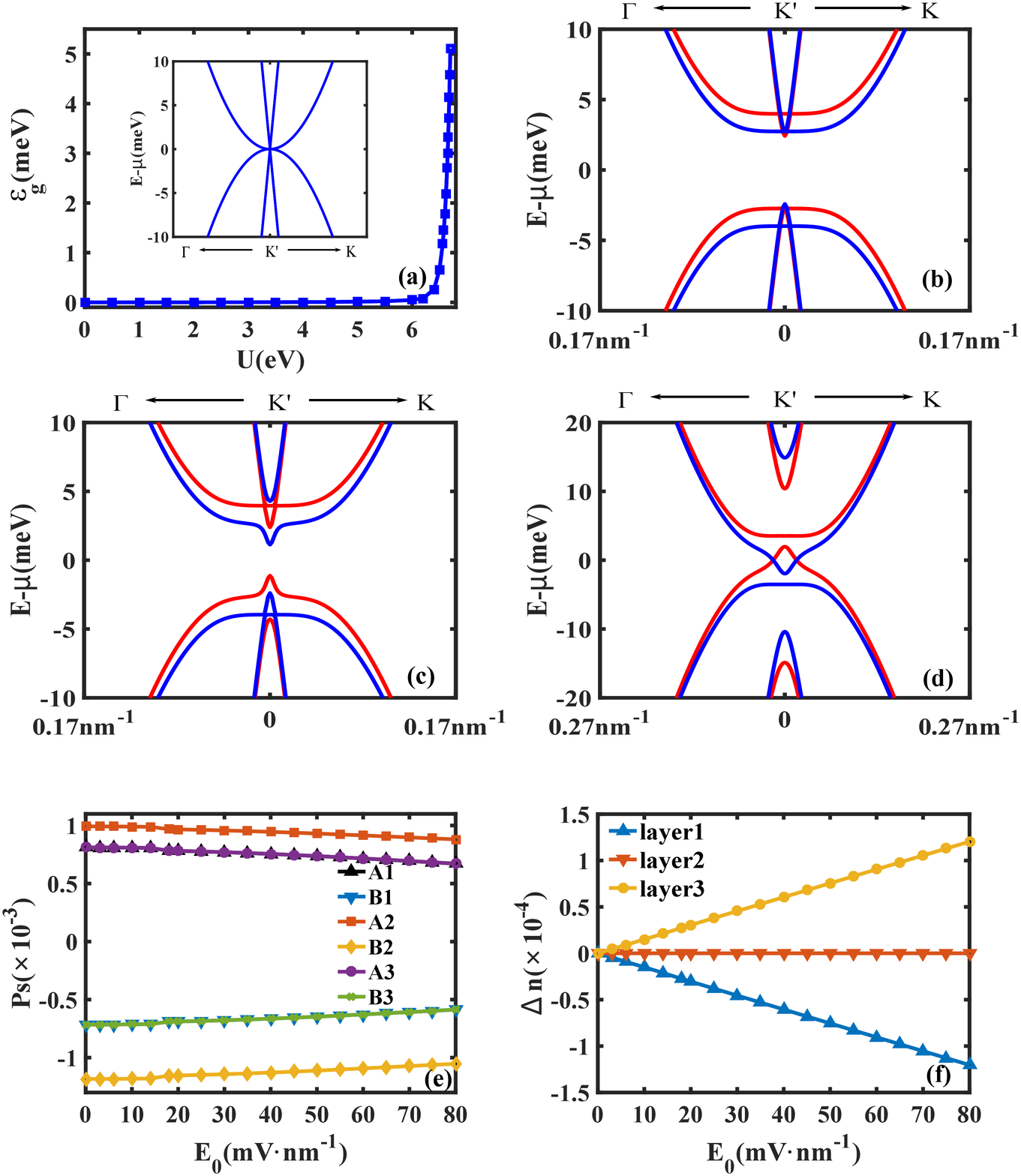}
\caption{(Color online) 3L-BGM at half filling. (a) Energy gap of 3L-BGM as a function of $U$ and in the absence of $E_0$. Inset shows the low energy bands with $U=0$. (b) Low energy bands with parameters $U=6.70$ $eV$ , $t=3.16$ $eV$, and $t_{\bot}=0.381$ $eV$. The following (c) and (d) use the same parameters of (b). (c) and (d) show the energy bands of the LAF states at the electric field (c) $E_0=10$ $mVnm^{-1}$ (d) $E_0=80$ $mVnm^{-1}$. (e) Spin polarization $P_s$ at each site in the unit cell as a function $E_0$. (f) The charge variation $\Delta n$ on each layer versus $E_0$.
}
\label{fig5}
\end{figure*}

In above, we consider the half metal phase in 4L-BGM. Other even-layer BGMs have similar single particle band structures, \textit{i.e.}, parabolic bands touching at the Fermi level. Meanwhile, interaction induced  correlated  ground state has also been observed in the Bernal stacked graphene hexalayer (\textit{i.e.}, 6L-BGM) in experiment\cite{2018A}. Thus, it is natural to expect that 6L-BGM can also have a half metallic state like that in 4L-BGM. We calculate the case of 6L-BGM as well, and the results are quite like the case of 4L-BGM. 6L-BGM does have a half metallic phase in the presence of a vertical electric field and tiny doping. And, the similar electric field induced spin-polarized band inversion has also been observed.   We argue that such correlated half metallic state is a general phenomenon of the even-layer BGM, as long as it has a LAF-like ground state. 

\begin{figure*}[ht]
\centering
\includegraphics[width=14cm]{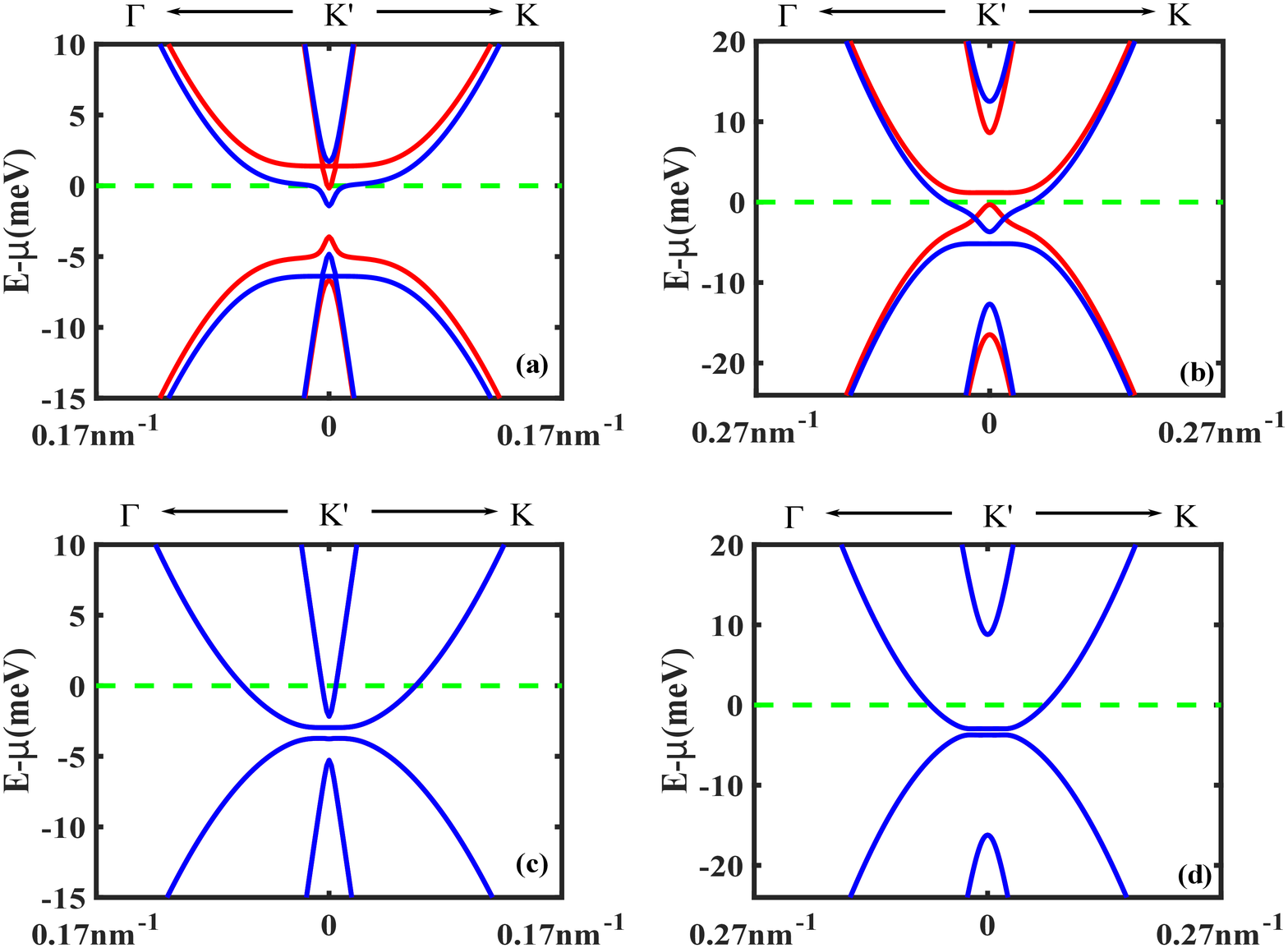}
\caption{(Color online) Half metallic phases in 3L-BGM. (a) and (b) are the energy bands of the proposed half metal states with electron density (a) $\delta n\approx0.34\times10^{10} cm^{-2}$ and (b) $\delta n\approx3.22\times10^{10} cm^{-2}$. (c) and (d) show the energy bands of the LCP states with a large charge density (c) $\delta n\approx1.29\times10^{11} cm^{-2}$ and (d) $\delta n\approx1.26\times10^{11} cm^{-2}$. The parameters used in (a) and (c) are the same as that in Fig.~\ref{fig5} (c), and the parameters in (b) and (d) are the same as that in Fig.~\ref{fig5} (d).
}
\label{fig6}
\end{figure*}

\subsection{Half metallic state in 3L-BGM}
Former DFT calculations\cite{dft2020} reported that the odd-layer BGM, like graphene trilayer (\textit{i.e.}, 3L-BGM), also can host a stable LAF state as illustrated in Fig.~\ref{fig4}, though its conductance is always finite (pseudogapped)\cite{2018A}.  Here, we calculate the electric field effects on the LAF state in 3L-BGM, especially the spin-resolved band structures.  The results are summarized in Fig.~\ref{fig5} and Fig.~\ref{fig6}. 
Our main findings are: (1)  First of all, the spin-up and spin-down bands of the LAF state in the 3L-BGM are no longer degenerate in the zero electric field case. (2) The influence of the vertical electric field on the LAF state in 3L-BGM is totally different from that in 4L-BGM. (3) The calculated band structures indicate that the 
required parameter range to  realize a half metallic phase is much smaller than that in 4L-BGM,  which implies that it is rather hard to observe the half metallic phase in 3L-BGM.

In single particle picture, 3L-BGM has a pair of parabolic bands and a pair of linear bands  touching at the the Dirac point, as shown in the inset of Fig.~\ref{fig5} (a). Coulomb interaction is able to induce a LAF state, and the corresponding spin-resolved band structure is given in Fig.~\ref{fig5} (b). We see that both the linear bands and the parabolic bands are gapped, and the spin structure of the LAF state is shown in Fig.~\ref{fig4} (a). 
The calculated LAF state here is qualitatively in agreement with the DFT calculation.  A notable difference is that, as shown in the DFT calculation,  the remote hopping will slightly shift the relative position of the  linear bands (about 10 meV), which is not included in our model.  The shifted  linear bands will always contribute a small but finite conductance in the transport measurements. And the pseudo gap observed in experiment should be the gap between the parabolic bands\cite{2018A}.  
However, we argue that our model here captures the main characteristics of the LAF state in 3L-BGM, \textit{i.e.}, the antiferromagnetic spin order and the interaction induced gap. The  induced gap $\epsilon_g$, as well as the pseudo gap,  relies on the $U$, which is shown in Fig.~\ref{fig5} (a).  Here, we set $U=6.70$ eV in order to fit the observed pseudo gap in experiment about $5$ meV\cite{2018A}. 

A remarkable phenomenon in 3L-BGM is that, in zero electric field, the spin-up and spin-down bands are no longer degenerate [see Fig.~\ref{fig5} (b)], where the valence and conduction bands have opposite spin polarization. It is quite different from the case of 4L-BGM, and  actually results from  the special lattice structure of 3L-BGM. In Fig.~\ref{fig4}, we see that 3L-BGM has an antiferromagnetic spin order as well, where spin-up (spin-down) electrons are mainly on sublattice A (B). Note that sublattice A and sublattice B are inequivalent here. Thus, from the point of the mean field, spin up and down electrons feel different potentials, which is different from that in 4L-BGM. This is the reason why the spin degeneracy of the bands are lifted in 3L-BGM. This interesting phenomenon further implies that, with two inequivalent sublattices, Hubbard model can intrinsically give rise to an antiferromagnetic state with  spin nondegenerate bands without need of an additional potential  to further break the spin  degeneracy of the bands, unlike that in ionic Hubbard model\cite{Garg_2014}. This prediction can be immediately tested in some simple lattice models, \textit{e.g.}, Lieb lattice.

We then discuss the influence of the vertical electric field $E_0$. 
In Fig.~\ref{fig5} (c), we apply a small electric field $E_0=10$ $mVnm^{-1}$. Compared with the zero field case in Fig.~\ref{fig5} (b), a small $E_0$ only slightly  modifies the shape of the spin-resolved bands, and moves the conduction (valence) band downwards (upwards).
Note that $E_0$ does not evidently enhance the split between the spin-up and spin-down bands.  The effect of $E_0$ is the same in the large electric field case, as shown in Fig.~\ref{fig5} (d) with $E_0=80$ $mVnm^{-1}$. Here, the valence band (spin-up) is shifted upwards  and the conduction band (spin-down) is moved downwards further, which gives rise to a negative gap at the Fermi level. Similarly, the split between the spin-up and spin-down bands is still very small. Further calculations indicate that the LAF state in 3L-BGM is very robust against $E_0$, where the antiferromagnetic spin order still exist in an extremely large $E_0=2$ $eVnm^{-1}$. As functions of $E_0$, the spin polarization of each sites  is plotted in Fig.~\ref{fig5} (e) and the corresponding spin structure is illustrated in Fig.~\ref{fig4} (b). The charge variation of each layer is given  in Fig.~\ref{fig5} (f). We see that the change of spin polarization is very slow when increasing $E_0$. Meanwhile, as $E_0$ increases, charges are mainly transferred from the bottom layer (layer 1) to the top layer (layer 3), while that in the middle layer is nearly invariant. 

Though the spin split of the bands is not very large in 3L-BGM, a proper doping can still induce a half metallic phase in principle.
In Fig.~\ref{fig6}, we show the induced half metallic phases in 3L-BGM. In Fig.~\ref{fig6} (a), we apply an electric field $E_0=10$ $meVnm^{-1}$ and a doping $\delta n =0.34 \times 10^{10} cm^{-2}$. The Fermi level is shifted into the conduction band (spin-down), so that the spin-down electrons are conducting and the spin-up electrons are still gapped. It thus gives rise to a half metal. In Fig.~\ref{fig6} (b),  we apply an electric field $E_0=80$ $meVnm^{-1}$ and a doping $\delta n =3.22 \times 10^{10} cm^{-2}$. 
In this case, only spin-down electrons are conducting as well, so that it is also a half metal. However, we note that, in both cases, the spin-up bands are very close to the Fermi level. Thus, in reality, it is hard to detect the predicted spin-dependent gap in transport experiment. In other words, 3L-BGM is not a good platform to realize the half metallic state.  The detectable feature of the half metal in 3L-BGM  may be the nonzero net magnetic moment in the presence of electric field and doping. Since the required doping is of the order of $10^{-10} cm^{-2}$, the doping induced magnetic moment should be of the order of $10^{-10} \mu_B cm^{-2}$. Note that a larger doping will directly kill the LAF states of 3L-BGM, see Fig. \ref{fig6} (c) and (d), where the spin-up and spin-down bands become degenerate.  

\section{Summary}
We theoretically propose that it is possible to produce a half metallic phase in the BGM systems by applying a vertical electric field and a proper doping.  
Based on  Hubbard model and self-consistent method, we systematically study the correlated ground states of 4L-BGM and 3L-BGM in the presence of a vertical electric field and doping, where 4L-BGM (3L-BGM) is considered as an typical example of even-layer (odd-layer) BGM.  Our results show that both the 4L-BGM and 3L-BGM can support a stable LAF state with antiferromagnetic spin order. In 4L-BGM, a vertical electric field will  lift the degeneracy of the spin-up and spin-down bands, and gives rise to a spin-dependent gap. Increasing $E_0$, we find an interesting phenomenon of spin-polarized band inversion, where the spin polarization of the conduction and valence bands are inverted. It is an unique feature of the LAF state in even-layer BGM. Furthermore,  A tiny doping can shift the Fermi level into the spin-polarized conduction band (or valence band), which thus give rise to a half metallic phase with nonzero net magnetic moment.  The  3L-BGM has different behaviors. Due to its special lattice structure, the spin degeneracy of the bands is lifted even in the zero electric field case. The conduction and valence bands have opposite spin polarization here. As $E_0$ increases, the gap between the conduction and valence bands decreases and a negative gap can be achieved. However, the split between the spin-up and spin-down bands is always quite small no matter whether $E_0$ is applied or not. Thus, though doping can still induce a half metallic phase like that in 4L-BGM, we expect that it is hard to detect the half metallic phase in  3L-BGM. We note that the antiferromagnetic spin order (LAF state) in 3L-BGM is very robust against $E_0$.

From the point of experiment, we suggest that 4L-BGM is an ideal platform to detect the proposed half metallic phase in BGMs. The feasible ways may be to measure the spin-resolved gap or the nonzero net magnetic moment.  Finally, we would like to mention that, in two dimensional (2D) antiferromagnetic systems, vertical electric field induced half metallic state should be a general phenomenon, which is found not only in graphene systems but also in other 2D antiferromagnets\cite{Gong8511,Zeng_2020, Lv_2021}.

\begin{acknowledgments}
We acknowledge support from the National Natural Science Foundation of China (Grants No.11874160, 11534001), the National Key Research and Development Program of China (2017YFA0403501), and the Fundamental Research Funds for the Central Universities (HUST: 2017KFYXJJ027). 
\end{acknowledgments}

\bibliography{half_metal}

\end{document}